# The Mechanism for the Energy Buildup Driving Solar Eruptive Events

K. J. Knizhnik[*], S. K. Antiochos, C. R. DeVore, and P. F. Wyper[†]

Heliophysics Science Division, NASA Goddard Space Flight Center, Greenbelt, MD 20771, USA.

*Correspondence to: kalman.knizhnik.ctr@nrl.navy.mil. Current address: Space Science Division, Code 7683, Naval Research Laboratory, Washington, DC 20375-5337

†Current address: Department of Mathematical Sciences, Durham University, Stockton Road, Durham, DH1 3LE, UK.

# ABSTRACT

The underlying origin of solar eruptive events (SEEs), ranging from giant coronal mass ejections to small coronal-hole jets, is that the lowest-lying magnetic flux in the Sun's corona undergoes the continual buildup of stress and free energy. This magnetic stress has long been observed as the phenomenon of "filament channels:" strongly sheared magnetic field localized around photospheric polarity inversion lines. However, the mechanism for the stress buildup – the formation of filament channels – is still debated. We present magnetohydrodynamic simulations of a coronal volume that is driven by transient, cellular boundary flows designed to model the processes by which the photosphere drives the corona. The key feature of our simulations is that they accurately preserve magnetic helicity, the topological quantity that is conserved even in the presence of ubiquitous magnetic reconnection. Although small-scale random stress is injected everywhere at the photosphere, driving stochastic reconnection throughout the corona, the net result of the magnetic evolution is a coherent shearing of the lowest-lying field lines. This highly counter-intuitive result – magnetic stress builds up locally rather than spreading out to attain a minimum energy state – explains the formation of filament channels and is the fundamental mechanism underlying SEEs. Furthermore, this mechanism may be relevant to other astrophysical or laboratory plasmas.

# 1. Introduction

The solar corona has long been observed to exhibit frequent ejections of matter and magnetic field into interplanetary space, usually accompanied by bursts of ultraviolet (UV) to X-ray radiation. These solar eruptive events (SEEs) range in energy from $10^{32}$ ergs or more for a major coronal mass ejection (CME)/eruptive flare to $10^{27}$ ergs or less for a coronal-hole jet (e.g., Raouafi et al. 2016). Given the short time scale for all these events – tens of minutes for large CMEs/flares to minutes for jets – SEEs are widely believed to be due to the explosive release of magnetic energy stored in the corona, specifically in the free energy of the highly stressed fields of filament channels (FCs). These structures have long been postulated to be the energy source for large SEEs such as CMEs/eruptive flares (e.g., Forbes 2000; Mackay et al. 2010), but recent observations indicate that even small jet-like eruptions are due to the release of magnetic energy stored in mini-filaments (Sterling et al. 2015; Kumar et al. 2017). Consequently, understanding the mechanism for the formation of FCs is equivalent to understanding how the free energy for SEEs builds up in the corona, one of the most important problems in heliophysics.

FCs consist of strongly sheared magnetic flux that is highly localized about a polarity inversion line (PIL; e.g., Mackay 2015). They are ubiquitous throughout the corona, occurring along all types of PILs, from the strongest active regions to the quietest high-latitude regions bordering polar coronal holes. In fact, it is generally believed that a FC will eventually develop at every large-scale PIL (Mackay 2015). Cool (~ $10^4$ K) material frequently (but not always) collects in FCs to form the observable filaments and prominences that have been studied for centuries. Numerous observations, including measurement of the vector magnetic field at the photosphere (e.g., Rust 1967; Leroy et al. 1983; Martin 1998) and direct measurement of the vector field in prominences (Casini et al. 2003; Kuckein et al. 2012), have established definitively that field lines in FCs are long and greatly stretched out parallel to the PIL, rather than short and perpendicular. These long "sheared" magnetic field lines are also mandated theoretically in order to support the dense prominence material against gravity (Tandberg-Hanssen 1995). An important result is that FCs are the only places where the corona is strongly non-potential; the rest of the closed-field corona is typically observed to consist of smooth and laminar, quasi-potential loops (e.g., Schrijver et al. 1999).

Given the one-to-one association between FCs and PILs, the two classic models for the formation of FCs invoke the property unique to PILs that flux emerges and cancels there. In the emergence model, magnetic shear forms along a PIL as a result of the large-scale emergence of a subsurface twisted flux rope. The basic idea is that a horizontal twisted flux rope can emerge only partially; the concave-up portion below the central axis stays trapped below the photosphere by the weight of the dense plasma. In this case, the outer highly twisted flux of the rope appears in the corona as a high-lying quasi-potential arcade, while the inner axial rope flux appears as low-lying shear field running almost parallel to the PIL. Numerical simulations have confirmed that emergence generally leads to the formation of sheared flux localized along a PIL, similar to observed FCs (e.g., Manchester 2001; Fan 2001; Magara & Longcope 2003; Leake et al. 2013; Archontis et al. 2013). Large-scale flux emergence occurs only in young active regions, however, whereas FCs form on all PILs, even those where no flux is emerging. The recent jet observations demonstrate, in particular, that mini-filaments are not due to flux emergence (Sterling et al. 2015, Kumar et al. 2017). Consequently, emergence cannot explain the majority of FCs.

In the cancellation model, the basic idea is that the coronal magnetic field acquires large-scale shear due to differential rotation or other large-scale motions, and/or to active-region emergence. This shear convects with the field toward PILs, where positive and negative fluxes undergo reconnection at the photosphere and submerge while the shear remains in the corona and piles up at the PILs (van Ballegooijen & Martens 1989; Mackay et al. 2010). The drawback with the cancellation theory is that outside of filaments, the coronal magnetic field exhibits only small stress. Many high-resolution observations (e.g., Schrijver et al. 1999) have shown that coronal loops are laminar and not far from their current-free state, in agreement with theoretical expectations (e.g., Parker 1983). Both theory and simulations demonstrate, however, that when low-shear field lines reconnect systematically along a line, the resulting field lines are highly twisted (van Ballegooijen & Martens 1989; Mackay et al. 2010). This process is exactly as occurs in the well-known two-ribbon flares, where systematic reconnection of coronal flux along an X-line produces the highly twisted flux rope observed as an ICME (e.g., Longcope 1996; Qiu et al. 2007; Gopalswamy et al. 2017). Strong twist, however, is not observed in filaments (e.g., Lin et al. 2005; Vourlidas et al. 2010) or anywhere else in the corona, except as the aftermath of flare reconnection. In order to explain observed FCs, a mechanism is needed that produces strong shear at *all* PILs, but with negligible twist.

A model that may satisfy these challenging requirements has been proposed recently by Antiochos (2013). This new "helicity condensation" model builds upon the standard processes postulated for the quasi-steady heating of the corona (e.g., Klimchuk 2006). First, magnetic stress is injected into the coronal magnetic field by the turbulent cellular convective motions and the continual emergence and submergence of small-scale flux throughout the photosphere, the so-called magnetic carpet (Title 2000). Reconnection then releases much of the free energy in this small-scale stress, heating the plasma and keeping the coronal field outside of filaments laminar, as observed (Schrijver et al. 1999).

The helicity condensation theory adds to these well-known processes the key concept of magnetic helicity conservation. Magnetic helicity is the topological measure of field-line linkages (e.g., Moffatt & Ricca 1992; Berger & Field 1984; Finn & Antonsen 1985). The all-important property of helicity is that it is conserved under magnetic reconnection (Woltjer 1958) and, therefore, provides strict constraints on the evolution and equilibrium state of the coronal magnetic field (e.g., Taylor 1974, 1986). If the stress injected into the corona contains net magnetic helicity, this helicity cannot be destroyed by reconnection and, therefore, must build up. There is, in fact, compelling evidence for such net helicity injection. Many observations have shown that in the Sun's northern hemisphere, large-scale structures such as active regions, filaments, and supergranular flows predominantly have negative helicity, whereas those in the southern have positive (Martin et al. 1992; Rust & Kumar 1994; Zirker et al. 1997; Pevtsov et al. 2003). A recent detailed analysis of erupting filaments indicates that this helicity preference may be as large as 90% (Ouyang et al. 2017). The helicity condensation theory argues that if the hemispheric preference extends down to the scales of the convective flows and flux emergence, then the helicity associated with the small-scale stress injected into the corona will be transformed by coronal reconnection into large-scale magnetic shear localized about PILs (Antiochos 2013).

This conjecture has been tested numerically (Zhao 2015; Knizhnik et al. 2015, 2017), but only within a simplified Parker (1972) coronal model, which consists of an initially uniform vertical field between two horizontal planes that represent the photosphere. Small-scale flows on these planes were used to inject free energy and helicity into the system. Although the Parker model is highly useful for studying coronal heating, its major shortcoming for testing helicity condensation is that there is no PIL. The simulations instead divided the domain into an inner region with photospheric driving flows and an outer undriven region, with the boundary between them assumed to represent the PIL. As a result of photospheric driving and coronal reconnection, the magnetic stress did accumulate at this "PIL;" however, such an evolution is also expected from energy arguments. Simple energy minimization would demand that any net magnetic stress should spread out as much as possible into the undriven region. Although the Parker-model simulations provide important insight into the physical process of helicity condensation, they cannot definitively account for the formation of observed FCs and, hence, cannot explain the energy buildup that is responsible for SEEs. We describe below the results of the first rigorous study of the helicity condensation model in a realistic coronal system with a true PIL and an open-flux coronal-hole region.

## 2. Numerical Model

For these calculations we used the Adaptively Refined MHD Solver (ARMS; DeVore & Antiochos 2008) to solve the standard ideal MHD equations in Cartesian coordinates in a rectangular domain. As is usual for such simulations, magnetic reconnection occurs due to numerical diffusion at locations where the current density develops structure at the grid scale. The domain size is $[0,L_x]\times[-L_y,+L_y]\times[-L_z,+L_z]$, where $x$ is the vertical direction (normal to the photosphere), $L_x = 2.0$, and $L_y = L_z = 3.5$. We employ zero-gradient conditions at all times at all six boundaries,

$$\frac{\partial \rho}{\partial n} = 0, \quad \frac{\partial T}{\partial n} = 0, \quad \frac{\partial \boldsymbol{v}}{\partial n} = 0, \quad \frac{\partial \boldsymbol{B}}{\partial n} = 0,$$

where $n = x,y,z$ is the normal coordinate. With the exception of the bottom boundary, all boundaries are open and free slip, so that both plasma and magnetic field can move along and across each boundary. The bottom boundary is closed and the field is line tied, so that the field lines move only in response to the imposed boundary flows described below. This emulates the slow driving at the dense solar photosphere, overriding the response to coronal magnetic and other forces.

The initial plasma parameters used in our dimensionless simulations are $\rho_0 = 1$, $T_0 = 1$, $P_0 = 1$, so that the sound speed is $c_s = (\gamma P_0/\rho_0)^{1/2} = 1.3$ with $\gamma = 5/3$. From the initial magnetic field prescribed in the domain, described below, the initial Alfvén speed $c_{A0} = B_0/(4\pi\rho_0)^{1/2}$ is found to range from 1.4 to 21.2, and the initial plasma beta (ratio of plasma thermal pressure to magnetic pressure) $\beta_0 = 8\pi P_0/B_0^2$ is found to range from $4\times10^{-3}$ to 1. The $\beta \ll 1$ regime corresponds to a magnetically dominated plasma, resembling the corona. The regions where $\beta \approx 1$ are mostly in the upper corners of the domain, so the amount of high-beta plasma is minimal and is not found to affect the results.

The initial magnetic field superposes a uniform vertical background field, $B_0 = -4$, and a potential bipolar sunspot whose normal magnetic field on the photospheric boundary ($x = 0$) is

$$B_{ss}(r) = \frac{B_+}{2}\left\{1 - \tanh\left(\frac{r^2 - r_+^2}{\lambda_+^2}\right)\right\} - \frac{B_-}{2}\left\{1 - \tanh\left(\frac{r^2 - r_-^2}{\lambda_-^2}\right)\right\}.$$

Here $r = (y^2+z^2)^{1/2}$ is the cylindrical radial coordinate on the plane; $B_+ = 60$, $\lambda_+ = 0.5$, and $r_+ = 2$; and $B_- = 15$, $\lambda_- = 1$, and $r_- = 4$. The PIL is located just beyond $r = r_+ = 2$.

Figure 1 shows the simulation domain along with some representative initial field lines. The black and white field lines of Figure 1 are tied at both ends corresponding to the closed corona, whereas the yellow lines are open at the top ends corresponding to field lines opening into the solar wind. Although the flux distribution of Figure 1 is clearly far simpler than observed photospheric flux distributions, our system nevertheless captures the essential features of a bipolar field with closed and open field lines in the corona and a PIL on the photosphere.

To model the photospheric convection, we placed 199 convective cells on the bottom boundary, arranged in the manner shown in Figure 1. The flows fill the negative polarity region as much as possible, injecting magnetic stress into all the coronal flux except for a tiny band about the PIL. Applying flows at the PIL itself would result in the mixing of positive and negative flux, making it impossible to resolve numerically the magnetic evolution near the PIL. Even with the restricted driving, the numerical resolution near the PIL becomes marginal, resulting in the small (~10%) violation of helicity conservation discussed below (Figure 2). By design, the flows preserve the vertical flux distribution, $B_x$, on the surface. In the interior region where $B_x$ is essentially uniform, each individual cell has an angular rotation rate given by

$$\Omega(r,t) = \begin{cases} -\Omega_0 f(t) g(r) & r \leq a_0 \\ 0 & r \geq a_0 \end{cases}$$

where

$$f(t) = \frac{1}{2}\left[1 - \cos\left(2\pi \frac{t}{\tau}\right)\right],$$

$$g(r) = \left(\frac{r}{a_0}\right)^4 - \left(\frac{r}{a_0}\right)^8,$$

and here $r = ([y-y_0]^2+[z-z_0]^2)^{1/2}$ is the cylindrical radial coordinate relative to the cell center at $(y_0,z_0)$. We chose $a_0 = 0.125$, $\tau = 0.572$, and $\Omega_0 = 43.94$, which set the peak linear velocity $v_{max} = 1.2$, about 5% of the peak Alfvén speed. The flows turn on from zero, and then turn off back to zero, at times $t = m\tau$ for integer $m \geq 0$. In some cases, we used the pauses in the flow at these times either to rotate the entire pattern of convective cells about the center of the sunspot or to flip the sense of the rotation (clockwise or counter) according to a random probability assigned to each cell.

We calculated the magnetic helicity $H$ in the volume (Finn & Antonsen 1985) by evaluating numerically

$$H = \int_V (A + A_P) \cdot (B - B_P) dV,$$

where $A_P$ is the vector potential and $B_P$ the magnetic field for the initial current-free state given above, and $A$ and $B$ are the corresponding instantaneous vector potential and magnetic field of the evolving solution. We also calculated the rate of helicity injection from the time derivative of $H$ assuming that the magnetic field evolves ideally in the volume,

$$\frac{dH}{dt} = \int_S \{(A_p \cdot v)B - (A_p \cdot B)v\} \cdot dS,$$

where $S$ denotes the photospheric plane $x = 0$. The results of the surface and volume integrals are shown in Figure 2. The black curve is the time integral of the rate of helicity injection, $dH/dt$; the red curve is the volume helicity, $H$. After 90 cycles, there is only a 10% difference between the injected and accumulated helicity, verifying that the evolution shown below is dominated by reconnection rather than diffusion.

## 3. Results

In order to test the helicity-condensation mechanism for FC formation, we performed several simulations with varying assumptions on the driving flows. In the case shown below, each small-scale flow was imposed for a maximum rotation of $\pi$, then the whole flow pattern was shifted by a random angle, and a new rotation applied. This procedure was performed for 90 cycles of rotation, followed by 7.5 cycles with no driving to allow the system to relax. Due to the random shifting of the pattern, the driving motion is truly chaotic, as is the driving expected from the solar convective flows and the magnetic carpet (Duvall & Gizon 2000; Title 2000). We also investigated cases where the flow pattern was held fixed throughout the simulation, and found that the results were quantitatively similar to the corresponding random-driving cases. To maximize the rate of helicity injection and minimize the computational time required, all flows in the simulation shown below were assumed to have the same (clockwise) sense of rotation. We also investigated cases in which every small-scale flow was assigned a probability governing the sense of rotation for each cycle. For example, assigning a 50% probability of clockwise vs. counter-clockwise produces zero net helicity injection. This case resulted in no sheared-field formation at the PIL and no continuous free-energy buildup. Assigning a 75% clockwise probability and 25% counter leads to essentially identical results to the ones below, except that the rates of helicity and energy buildup were 50% smaller. These results demonstrate that the details of the flow pattern are not important for the FC formation process. The only requirements are that net helicity be injected into the corona and that reconnection occurs quasi-chaotically throughout the corona.

Figure 3 shows the end state of our randomly shifted system (the detailed qualitative evolution can be seen in the accompanying movie). This result is an amazing example of self-organization. First, note that the yellow open lines are almost unchanged from their initial state in Figure 1. This is expected, because any stress injected onto these lines simply propagates out of the system. The white field lines near the center of the closed field region are somewhat changed from their initial state, but far less than would be expected for an ideal evolution. If there were no coronal reconnection, the footpoint driving for this random case would produce a completely tangled, twisted mess; the fixed-flow case, in contrast, would produce a set of parallel, highly twisted (~ 90 rotations), small flux tubes. In both cases, we find instead that the white field lines are smooth and laminar – similar to their initial potential state – except that they bulge upward substantially, appearing very much like the coronal loops that are commonly observed in the closed corona (Schrijver et al. 1999). Our results demonstrate that small-scale, "chaotic" reconnection can explain these observations, in spite of the complex driving from photospheric convection and small-scale flux emergence.

The black field lines near the PIL, on the other hand, are drastically changed from their initial state. They are now greatly stretched out along the PIL, with the ratio of parallel to perpendicular field component ~ 20. The black field lines have exactly the properties required to account for FCs. They are purely sheared, roughly parallel, with negligible internal twist, *even though all the driving motions were purely rotational with no large-scale shear*. This is a critically important point. The results shown are due solely to the effects of coronal reconnection and helicity conservation and, therefore, are completely general. The same structure would result for any form of small-scale complex driving, as long as the driving injects net helicity into the system. We expect, therefore, that FCs and the resulting energy buildup leading to eruptions will occur in any corona-like system, such as the atmospheres of late-type stars, accretion disks, etc.

To show that the state above is ripe for magnetic eruption, we plot in Figure 4 the shear flux through a vertical plane together with some field lines. The low-lying flux bulges outward in order to minimize the magnetic pressure associated with its strong shear, but it is held back by the magnetic tension of the overlying unsheared flux. This is exactly the force balance believed to underlie all CMEs/flares (Priest 2014) and even coronal jets (Wyper et al. 2017).

It is important to understand physically why the magnetic state of Figures 3 and 4 develops. Obviously, it is not due to energy minimization. For a "turbulently" reconnecting system in which helicity conservation is the only constraint, the minimum energy state is a linear force-free field (FFF) (e.g., Taylor 1974; Heyvaerts & Priest 1984). The system above is close to force-free, but the stresses (i.e., the electric currents) are tightly concentrated near the PIL rather than distributed uniformly throughout the field as in a linear FFF. In fact, the only linear FFF appropriate to our system is the potential (current-free) state of Figure 1, because the open flux cannot support steady electric currents. Clearly, our system has additional constraints beyond helicity conservation, most likely due to the photospheric line-tying (e.g., Antiochos et al. 2002).

The results of our calculations can be understood, instead, as due to the well-known inverse cascade of helicity in turbulence theory, but with a curious twist. An inverse cascade implies that helicity should evolve so as to "condense" at the largest scale of the system, its boundary (Biskamp 1993). Normally, this boundary coincides with the largest spatial scale. The system of Figure 1 has two boundaries that define the amount of closed flux: an inner circle separating the open (yellow) and closed (white) field lines, and the blue PIL. The former corresponds to the largest spatial scale, in that the longest field lines occur there. However, any helicity that condenses at this boundary simply propagates away as Alfvén waves along the infinitely long open lines, thereby minimizing the energy as well. Our striking result is that helicity also condenses onto the other boundary, the PIL, even though it corresponds initially to the shortest field lines in the system. Helicity condensation at this boundary continuously forms a filament channel, resulting in a constant buildup of free energy, which can be removed only through ejection. We conclude, therefore, that the highly counter-intuitive mechanism of helicity condensation may finally explain why the solar corona continually undergoes magnetic eruption.

## Acknowledgments

Our work was supported by NASA's LWS, H-SR, NESSF, and HEC programs.

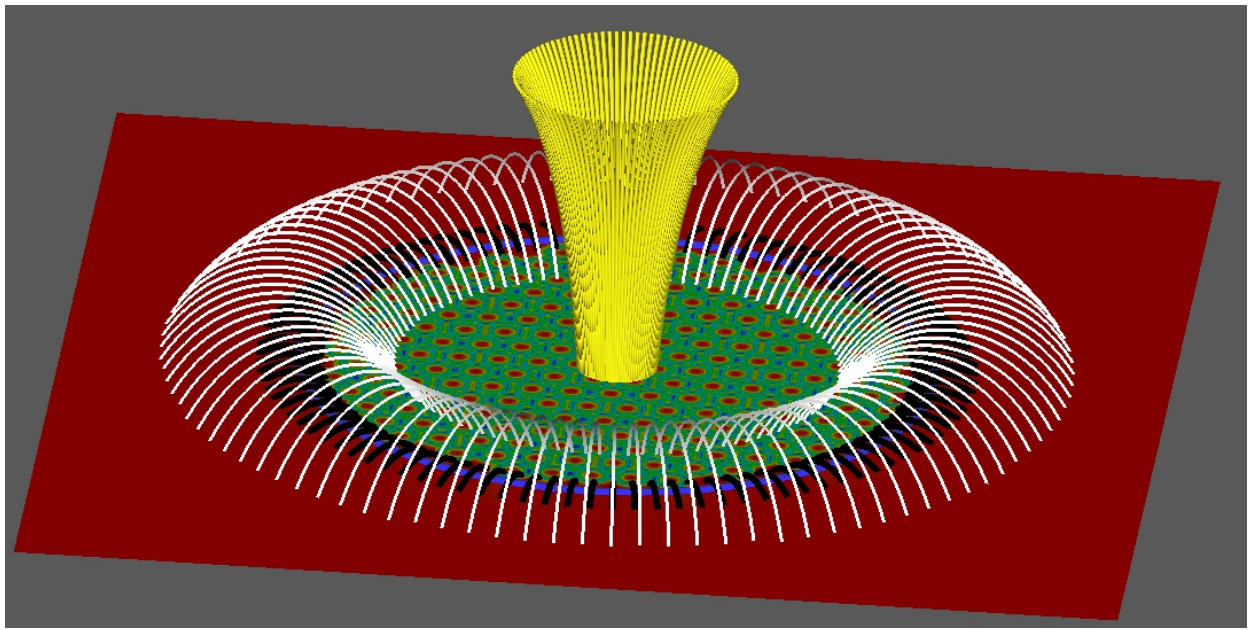

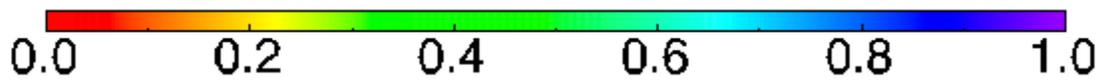

**Figure 1.** Simulation setup showing closed, line-tied field lines (white and black) traced from the bottom boundary crossing the circular PIL (blue), and open, free-slip field lines (yellow) traced from the top boundary. Bottom boundary shading shows velocity magnitude due to the driving inside the PIL.

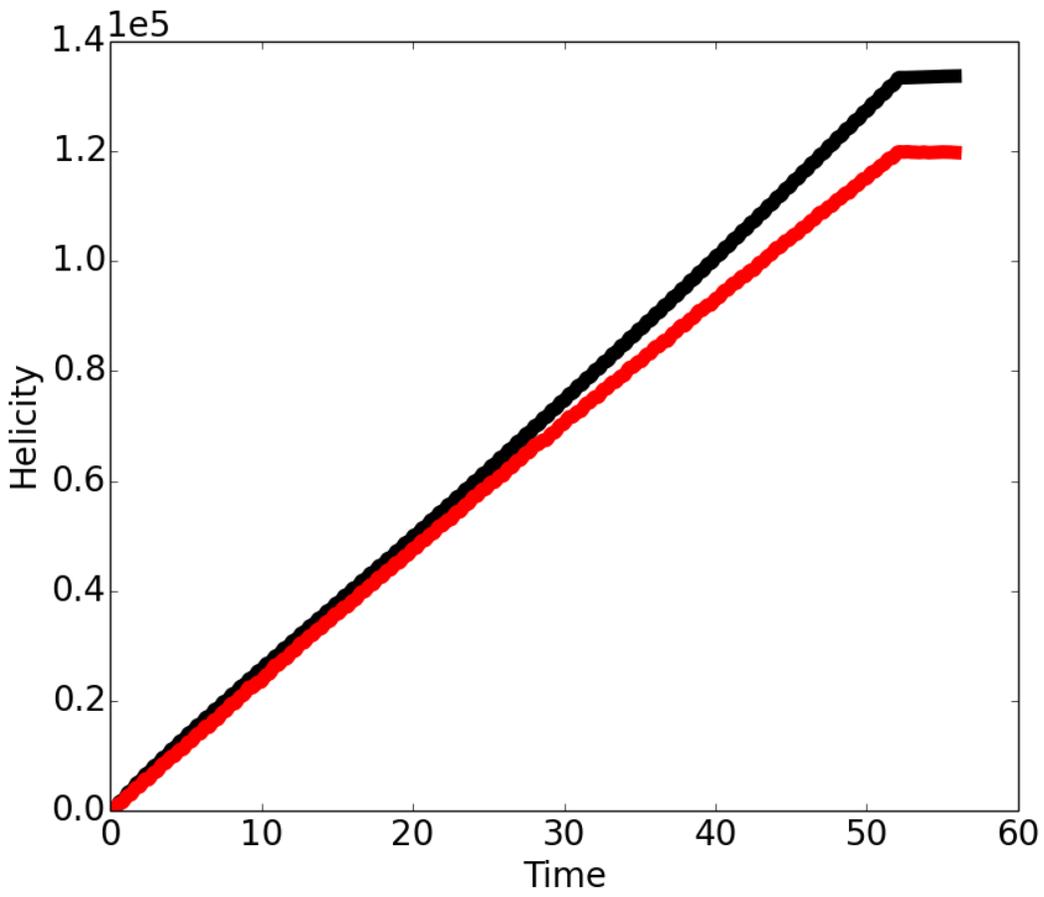

**Figure 2.** Accumulated magnetic helicity injected through the bottom boundary (black) and stored in the volume (red) versus time.

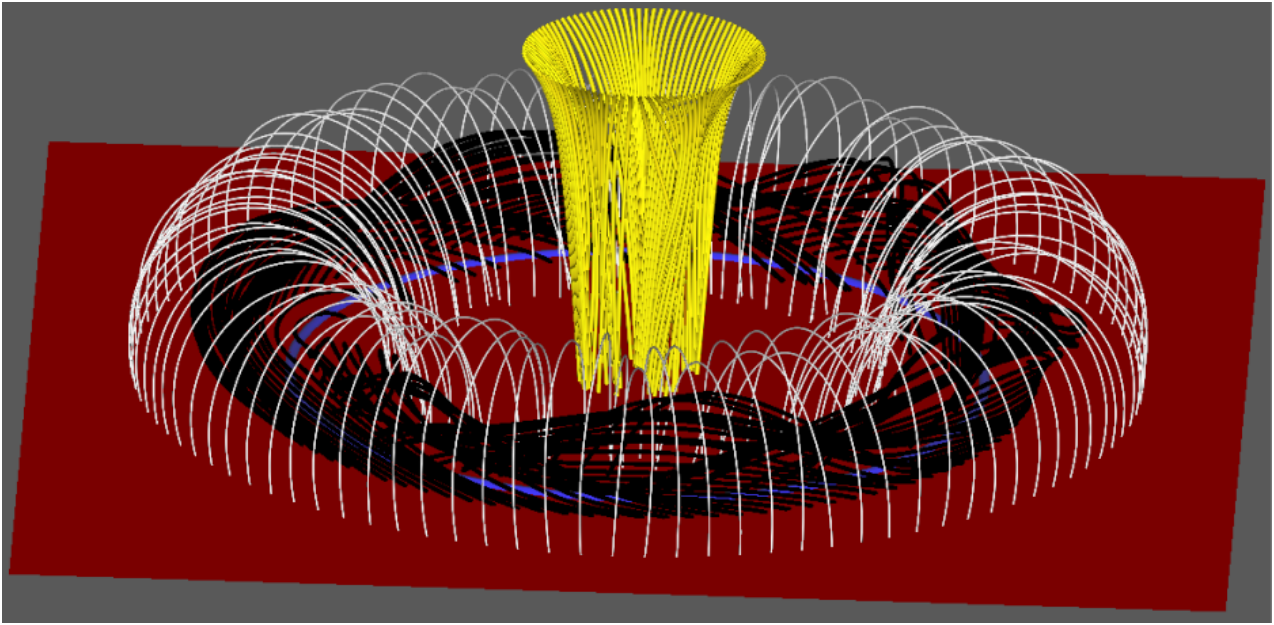

**Figure 3.** End state of Figure 1 after 90 randomly shifted rotation cycles and 7.5 relaxation cycles.

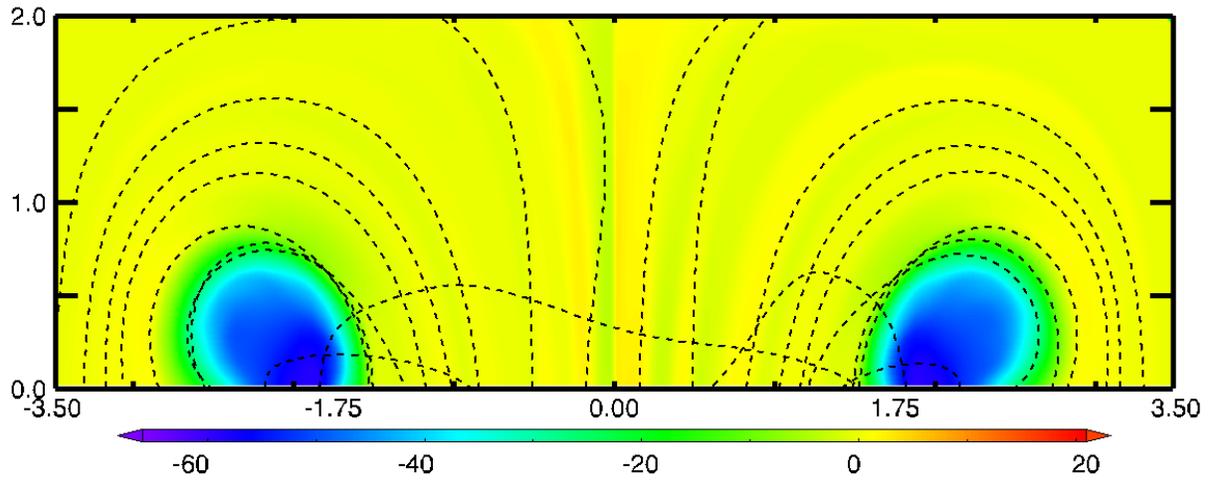

**Figure 4.** Azimuthal field (color shading) through a vertical plane between the top and bottom boundaries with field lines (dashed lines) projected onto the plane, at the same time as Figure 3.